\documentclass[prc,aps,showpacs,showkeys,groupedaddress,floatfix,superscriptaddress]{revtex4-1}
\usepackage{eurosym}
\usepackage{natbib}
\usepackage{epsfig}
\usepackage{dcolumn}
\usepackage{bm}
\usepackage[utf8]{inputenc}
\usepackage{graphicx}
\usepackage{amsmath}
\usepackage{amsfonts}
\usepackage{amssymb}
\usepackage{color}
\usepackage{float}

\begin{document}

\title{The spin-one DKP Equation with a nonminimal vector interaction in the
presence of minimal uncertainty in momentum }

\author{B. Hamil}
\email{hamilbilel@gmail.com}
\affiliation{D\'{e}partement de TC de SNV, Universit\'{e} Hassiba Benbouali, Chlef,
Algeria}

\author{B. C. L\"{u}tf\"{u}o\u{g}lu}
\email{bclutfuoglu@akdeniz.edu.tr}
\affiliation{Department of Physics, Akdeniz University, Campus 07058
Antalya, Turkey}
\affiliation{Department of Physics,  University of Hradec Kr\'{a}lov\'{e},
Rokitansk\'{e}ho 62, 500\,03 Hradec Kr\'{a}lov\'{e}, Czechia}

\author{H. Aounallah}
\email{houcine.aounallah@univ-tebessa.dz}
\affiliation{Department of Science and Technology. Larbi Tebessi University,
12000 Tebessa, Algeria.}
\date{\today}

\begin{abstract}
In this work, we consider the relativistic Duffin-Kemmer-Petiau equation for spin-one particles with a nonminimal vector interaction in the presence of minimal uncertainty in momentum. By using the position space representation we exactly determine the bound-states spectrum and the corresponding eigenfunctions. We discuss the effects of the deformation and nonminimal vector coupling parameters on the energy spectrum analytically and numerically.
\end{abstract}

\maketitle

\section{Introduction}

In a series of studies Duffin, Kemmer, and Petiau put forward a first order
differential equation, hereafter (DKPe), to describe the dynamics mesons
\cite{Petiau1936, Kemmer_1938, Kemmer_1939, PhysRev.54.1114}. Although the
DKPe presents similarities to the Dirac's equation, its matrices obey a
different and more complex commutation rule schemes. Until the 1970s, DKP
formalism developed \cite{Kinoshita_1950} with increasing doubts over the
DKPe's relationships with the Klein-Gordon (KG) and Proca equations \cite%
{Pimentel_2000}. Between 1970 and 1980, interest in the DKPe waned,
believing that DKPe was equivalent to the KG and Proca equations \cite%
{Boumali_et_al_2013}. After showing that the equivalence is valid only in
the special case where symmetry exists \cite{Krajcik_1977}, interest in the
solutions of the DKPe has increased in the last decades.

The DKPe with different types of couplings is used in a wide area of
physics. For example in the modelling of the: meson scattering by nuclei
\cite{CLARK1998231}, large and short distance interactions in quantum
chromodynamics \cite{Gribov_1999}, boson dynamics in curved space-time \cite%
{Castro_2015, Hun2019}, covariant Hamilton dynamics \cite{KANATCHIKOV2000107}%
, non-inertial effect of rotating frames \cite{Castro_2016}, Galilei
covariance \cite{Montigny_2000}, the Aharanov-Bohm phenomenon \cite%
{Houcine_et_al_2019, Boumali_et_al_2020}, dynamics of vector bosons in the
expanding universe \cite{Sucu2005}, commutative and noncommutative spaces
\cite{Boumali_et_al_2018, Falek_2008}, thermodynamic properties of bosons
\cite{ doi.org/10.1140/epjp/i2013-13124-y, Aounallah_et_al_2020},
Bose-Einstein condensation \cite{CASANA200333, ABREU2015612} and etc...

It is a well-known fact that the nonminimal vector couplings to the KG and
Proca equations produce results that contradict the predictions of the
non-relativistic quantum mechanics \cite{PhysRevD.15.1518,
Vijayalakshmi_1979, cardoso_estados_2008, DEOLIVEIRA2016320}. With a
nonminimal vector coupling, one refers to a sort of charge conjugate
invariant coupling which transforms like a vector under a Lorentz
transformation. If the nonminimal vector potential is invariant under charge
conjugation, then, one can not discriminate the particle from its
antiparticle \cite{Castro_et_al_2014}. Since the DKPe, unlike from KG and
Proca equations, allows the nonminimal couplings, it is extensively examined
by considering several Lorentz structures \cite{PhysRevC.40.2181,
PhysRevC.50.2624, AITTAHAR1995307, BARRETT1995311, doi:10.1063/1.3494292,
Cardoso_2010}, and potential energies \cite{doi:10.1142/S0217732312502288,
hassanabadi_duffinkemmerpetiau_2013, Molaee_2013, hhassanabadi_dkp_2013,
BAHAR2014105, doi:10.1142/S0218301314500074}.

On the other hand, various researches regarding the quantum gravity \cite%
{Veneziano_1986, AMATI198781} and cosmology \cite{Bosso_2020}, string theory
\cite{SCARDIGLI199939, Scardigli_2003}, noncommutative geometry \cite%
{MAGGIORE199383}, black hole physics \cite{XIANG2018334} and thermodynamics
\cite{Saghafi_2017} show that a minimal observable length should exist.This
minimal length (ML) may be introduced as an additional uncertainty in position
measurements $\Delta x_{\min }$, which leads to a generalization of
Heisenberg's uncertainty principle. Kempf with his collaborators \cite%
{doi:10.1063/1.530798, PhysRevD.52.1108, doi:10.1063/1.531501, Kempf_1997},
showed that a ML can be obtained out of  the generalized Heisenberg
algebra with the form%
\begin{equation}
\left[ X_{i},P_{j}\right] =i\hbar \delta _{ij}\left( 1+\alpha P^{2}\right) ;%
\text{ \ }\left[ X_{i},X_{j}\right] =i\hbar \alpha J_{ij},
\end{equation}
where $\alpha $ is the parameter of deformation. It is worth mentioning that the
deformed algebra leading to quantized space time was introduced for first
time by Snyder \cite{Snyder}.

It is well known that the curvature of space-time becomes important at great distances. On a general curved space-time, there is no concept of a plane wave.  This implies that there is a finite lower bound to the precision with which the corresponding momentum can be described. This can be represented with a nonzero minimal uncertainty in momentum (MUM) measurement. It has been argued in \cite{bolen_anti-sitter_2005} that in the presence of a
cosmological constant the Heisenberg uncertainty principle receives a
correction term due to the background curvature, which is known as the "extended
uncertainty principle" (EUP),%
\begin{equation}
\left( \Delta X_{i}\right) \left( \Delta P_{j}\right) \geq \frac{\hbar
\delta _{ij}}{2}\left( 1+\alpha \left( \Delta X_{i}\right) ^{2}\right) ,
\label{eup1}
\end{equation}%
where the deformation parameter is proportional to the cosmological constant. It is obvious that  Eq. \eqref{eup1} yields a nonzero minimal uncertainty in momentum  as $
\left( \Delta P\right) _{\min }=\frac{\hbar \sqrt{\alpha }}{2}$.  Recently, Mignemi showed that one can derive Eq. \eqref{eup1} from the definition of the quantum mechanics on the anti-de Sitter spacetime \cite{Mignemi2010}. Moreover, in that case, he stated that the modified Heisenberg algebra corresponding to the EUP as follows:
\begin{equation}
\left[ X_{i},P_{j}\right] =i\hbar \left( \delta _{ij}+\alpha
X_{i}X_{j}\right) ;\text{ }\left[ X_{i},X_{j}\right] =0;\text{ }\left[
P_{i},P_{j}\right] =i\hbar \alpha L_{ij}.  \label{eup2}
\end{equation}
Here, $L_{ij}$ is the angular momenta while $i,j=1,2,3$. In the position space, one particular explicit representation of the position and momentum operators that obeys Eq.~\eqref{eup2} is given with
\begin{subequations}
\label{ops}
\begin{eqnarray}
X_{i} &=&\frac{x_{i}}{\sqrt{1-\alpha r^{2}}},  \label{opx} \\
P_{i} &=&-i\hbar \sqrt{1-\alpha r^{2}}\frac{\partial }{\partial x_{i}};
\label{opp}
\end{eqnarray}
\end{subequations}
It is worth noting that as a consequence of the deformation of the usual algebra the conventional inner product definition is needed to be modified with \cite{Hamil_Merad_Birkandan_2020}
\begin{equation}
\langle \psi |\phi \rangle =\int \frac{d^{3}r}{\sqrt{1-\alpha r^{2}}}\psi
^{+}\phi .
\end{equation}

In the last decades by considering modifications to the momentum and position
operators the extension of the Heisenberg algebra is being examined
extensively  \cite{doi:10.1142/S0217732312500800,
PhysRevD.85.024016, doi.org/10.1140/epjp/i2013-13025-1, Ikot:2015kla,
doi.org/10.1140/epjp/i2017-11353-8, hamil_dirac_2018,
doi.org/10.1140/epjc/s10052-019-7463-3,
doi.org/10.1140/epjc/s10052-019-6718-3, MagrefId0,
Hamil_Merad_Birkandan_2020, BclEPL2020, farahani_dsr-gup_2020}.
However,  to the best of our knowledge, the rich structure of the DKPe due to non-minimum vector coupling has not been studied with a ML. Our purpose in this study is to consider the spin-one DKPe with a nonminimal vector interaction in the presence of the MUM.  The structure of the manuscript is constructed as follows: In section \ref{sec2} we briefly introduce the DKP formalism and discuss the nonminimal coupling vector interaction. In section \ref{sec3} we study the effect of the MUM on the spectrum of spin-one particles in the presence of a nonminimal vector linear potential in $(3 +1)$ dimensional. In the final section, we give our conclusion.


\section{The DKP Equation}

\label{sec2} In the case of non-interacting scalar and vector bosons the
DKPe is defined with the natural units, $(\hbar=c=1)$, as follows:
\begin{eqnarray}
\left[i\beta^\sigma\partial_\sigma-m\right]\Psi(\vec{r},t)=0, \qquad%
\mbox{where}\qquad \sigma=0,1,2,3.  \label{freeDKP}
\end{eqnarray}
Here, $m$ is the mass of the spin-one particle. The DKP matrices, $%
\beta^\sigma$, satisfy the following DKP algebra
\begin{eqnarray}
\beta^{\sigma}\beta^{\kappa}\beta^{\lambda}+\beta^{\lambda}\beta^{\kappa}%
\beta^{\sigma}=g^{\sigma\kappa}\beta^{\lambda}+g^{\kappa\lambda}\beta^{%
\sigma}.  \label{DKPalgebra}
\end{eqnarray}
where $g^{\sigma\kappa}=diag(1,-1,-1,-1)$ is the metric tensor of the
Minkowski space-time while $(g^{\sigma\kappa})^2=1$. In the spin-one sector,
the irreducible DKP matrices are given with $10 \times 10 $ matrix sets. In
this manuscript, we employ the following DKP matrices:
\begin{eqnarray}  \label{betamat}
\beta^{0}=\left(%
\begin{array}{cccc}
0 & \mathbf{\check{0}}_{1\times3} & \mathbf{\check{0}}_{1\times3} & \mathbf{%
\check{0}}_{1\times3} \\
\mathbf{\check{0}}^{T}_{3\times1} & \mathbf{\overline{0}}_{3\times3} &
\mathbf{I}_{3\times3} & \mathbf{\overline{0}}_{3\times3} \\
\mathbf{\check{0}}^{T}_{3\times1} & \mathbf{I}_{3\times3} & \mathbf{%
\overline{0}}_{3\times3} & \mathbf{\overline{0}}_{3\times3} \\
\mathbf{\check{0}}^{T}_{3\times1} & \mathbf{\overline{0}}_{3\times3} &
\mathbf{\overline{0}}_{3\times3} & \mathbf{\overline{0}}_{3\times3}%
\end{array}%
\right), \qquad \beta^{k}=\left(%
\begin{array}{cccc}
0 & \mathbf{\check{0}}_{1\times3} & \mathbf{u}^{j}_{1\times3} & \mathbf{%
\check{0}}_{1\times3} \\
\mathbf{\check{0}}^{T}_{3\times1} & \mathbf{\overline{0}}_{3\times3} &
\mathbf{\overline{0}}_{3\times3} & -i\mathbf{S}^{j}_{3\times3} \\
-\mathbf{u}^{j^{T}}_{3\times1} & \mathbf{\overline{0}}_{3\times3} & \mathbf{%
\overline{0}}_{3\times3} & \mathbf{\overline{0}}_{3\times3} \\
\mathbf{\check{0}}^{T}_{3\times1} & -i\mathbf{S}_{3\times3}^{j} & \mathbf{%
\overline{0}}_{3\times3} & \mathbf{\overline{0}}_{3\times3}%
\end{array}%
\right),
\end{eqnarray}
where $j=1,2,3 $,
\begin{equation}
\begin{array}{cc}
\mathbf{\check{0}}_{1 \times 3}=\left(%
\begin{array}{ccc}
0 & 0 & 0%
\end{array}%
\right), & \mathbf{I}_{3\times3}=%
\end{array}%
\left(%
\begin{array}{ccc}
1 & 0 & 0 \\
0 & 1 & 0 \\
0 & 0 & 1%
\end{array}%
\right),
\end{equation}%
\begin{equation}
\begin{array}{ccc}
\mathbf{u}^{1}_{1 \times 3}=\left(%
\begin{array}{ccc}
1 & 0 & 0%
\end{array}%
\right), & \mathbf{u}^{2}_{1 \times 3}=\left(%
\begin{array}{ccc}
0 & 1 & 0%
\end{array}%
\right), & \mathbf{u}^{3}_{1 \times 3}=\left(%
\begin{array}{ccc}
0 & 0 & 1%
\end{array}%
\right),%
\end{array}.%
\end{equation}%
$\mathbf{S}_{3\times3}^{j}$ are the usual spin-one matrices as given
\begin{equation}
\begin{array}{ccc}
\mathbf{S}^{1}_{3\times3}=i\left(%
\begin{array}{ccc}
0 & 0 & 0 \\
0 & 0 & -1 \\
0 & 1 & 0%
\end{array}%
\right), & \mathbf{S}^{2}_{3\times3}=i\left(%
\begin{array}{ccc}
0 & 0 & 1 \\
0 & 0 & 0 \\
-1 & 0 & 0%
\end{array}%
\right), & \mathbf{S}^{3}_{3\times3}=i\left(%
\begin{array}{ccc}
0 & -1 & 0 \\
1 & 0 & 0 \\
0 & 0 & 0%
\end{array}%
\right),%
\end{array}%
\end{equation}%
When the interactions are taken into account the DKPe is defined with
\begin{eqnarray}
\left[i\beta^\sigma\partial_\sigma-m-U\right]\Psi(\vec{r},t)=0, \qquad%
\mbox{where}\qquad \sigma=0,1,2,3.  \label{InteractingDKP}
\end{eqnarray}
with the natural units. Here, $U$ is the general potential energy matrix
that can be expressed with $100$ irreducible matrices in the spin-one
sector. In this case, the four-current, $J^\mu$, satisfies
\begin{eqnarray}
\partial_\mu J^\mu+ \frac{i\overline{\Psi}}{2}\left(U-\eta^0 U^\dagger
\eta^0\right)\Psi=0,
\end{eqnarray}
where $\overline{\Psi}=\Psi^\dagger \eta^0 $. Note that, the four-current is
conserved when $U$ is Hermitian with respect to $\eta^0$, \cite%
{Castro_et_al_2014}. In the spin-one sector the potential energy matrices
can be constructed by well-defined Lorentz structures such as two-vector,
two-scalar, two pseudo-vector, a pseudo-scalar, and eight tensor terms.
However, in applications tensor terms are discarded since they issue
non-causal effects \cite{Vijayalakshmi_1979}. In this manuscript we consider
a non-minimal vector interaction in the form of
\begin{eqnarray}
U&=&i \big[P,\beta^\mu \big]A_\mu.
\end{eqnarray}
Here, $P$ denotes the projection operator, thus, $P^2=P$ and $P^\dagger=P$.
It is worth noting that the considered potential energy matrices behaves as
a vector under the Lorentz transformation \cite{Castro_et_al_2014}. By
choosing the potential energy matrix in this way, it is shown that four
currents are conserved \cite{Castro_et_al_2014}.

In this manuscript, we consider a time-independent potential energy,
therefore we assume the spin-one wave function can be expressed in the form
of
\begin{equation*}
\Psi (\vec{r},t)=\psi (\vec{r})e^{-iEt}.
\end{equation*}%
Here, $E$ is the energy of the spin-one boson particle. Then, Eq. %
\eqref{InteractingDKP} reduces to
\begin{equation}
\left[ \beta ^{0}E+\overrightarrow{\beta }.\overrightarrow{P}-m-i\left[
P,\beta ^{\mu }\right] A_{\mu }\right] \psi (\vec{r})=0.  \label{eqA}
\end{equation}

\section{DKP equation in the presence of minimum uncertainty in momentum}

\label{sec3}

In this section, we examine the dynamics of a vector boson by solving the
DKPe in the presence of ML by considering a nonminimal vector coupling. We
take the wave function in the form of \cite%
{Hassanabadi_Molaee_Ghominejad_Zarrinkamar_2012}

\begin{equation}
\begin{array}{c}
\psi (\vec{r})=\left(
\begin{array}{c}
i\phi \\
\overrightarrow{F} \\
\overrightarrow{G} \\
\overrightarrow{H}%
\end{array}%
\right) ,%
\end{array}%
\end{equation}%
where $\phi $ is a scalar function, and
\begin{equation}
\begin{array}{ccc}
\overrightarrow{F}=\left(
\begin{array}{c}
\varphi _{2} \\
\varphi _{3} \\
\varphi _{4}%
\end{array}%
\right) , & \overrightarrow{G}=\left(
\begin{array}{c}
\varphi _{5} \\
\varphi _{6} \\
\varphi _{7}%
\end{array}%
\right) , & \overrightarrow{H}=\left(
\begin{array}{c}
\varphi _{8} \\
\varphi _{9} \\
\varphi _{10}%
\end{array}%
\right) .%
\end{array}%
\end{equation}%
We consider the parity operator as $P=\beta ^{\mu }\beta _{\mu }-2$. By
using the chosen representation, which is given in Eq. \eqref{betamat}, we
obtain the parity operator matrix as
\begin{equation}
P=diag\left(
\begin{array}{cccccccccc}
1 & 1 & 1 & 1 & 0 & 0 & 0 & 0 & 0 & 0%
\end{array}%
\right) .
\end{equation}%
We employ the position and momentum operators that are given in Eq. %
\eqref{ops} by considering the assumption of the presence of the ML. Then,
we derive a compact form of the time-independent DKPe out of Eq.
\eqref{eqA}
\begin{subequations}
\begin{eqnarray}
i\sqrt{1-\alpha r^{2}}\left( \overrightarrow{\nabla }-\frac{\overrightarrow{A%
}}{\sqrt{1-\alpha r^{2}}}\right) \times \overrightarrow{F} &=&m%
\overrightarrow{H},  \label{DKP1a} \\
\sqrt{1-\alpha r^{2}}\left( \overrightarrow{\nabla }+\frac{\overrightarrow{A}%
}{\sqrt{1-\alpha r^{2}}}\right) .\overrightarrow{G} &=&m\phi ,  \label{DKP1b}
\\
i\sqrt{1-\alpha r^{2}}\left( \overrightarrow{\nabla }+\frac{\overrightarrow{A%
}}{\sqrt{1-\alpha r^{2}}}\right) \times \overrightarrow{H} &=&m%
\overrightarrow{F}-\left( E-iA_{0}\right) \overrightarrow{G},  \label{DKP1c}
\\
\sqrt{1-\alpha r^{2}}\left( \overrightarrow{\nabla }-\frac{\overrightarrow{A}%
}{\sqrt{1-\alpha r^{2}}}\right) \phi &=&m\overrightarrow{G}-\left(
E+iA_{0}\right) \overrightarrow{F}.  \label{DKP1d}
\end{eqnarray}%
In order to solve these coupled equations we follow \cite%
{doi:10.1063/1.530801}, and assume that wave function components have the
form of
\end{subequations}
\begin{subequations}
\begin{eqnarray}
\phi &=&\frac{\varphi(r)}{r}Y_{JM}(\theta ,\phi )  \label{wf1a} \\
\overrightarrow{F} &=&\sum_{L}\frac{F_{nJL}(r)}{r}Y_{JL1}^{M}(\theta ,\phi )
\label{wf1b} \\
\overrightarrow{G} &=&\sum_{L}\frac{G_{nJL}(r)}{r}Y_{JL1}^{M}(\theta ,\phi )
\label{wf1c} \\
\overrightarrow{H} &=&\sum_{L}\frac{H_{nJL}(r)}{r}Y_{JL1}^{M}(\theta ,\phi ),
\label{wf1d}
\end{eqnarray}%
where $Y_{JM}(\theta ,\phi )$ is the spherical harmonics of order $J$, $%
Y_{JL1}^{M}(\theta ,\phi )$ are the vector spherical harmonics, and $\varPhi%
_{nJ}(r)$, $F_{nJL}(r)$, $G_{nJL}(r)$, and $H_{nJL}(r)$ are unnormalized
radial wave functions. In this manuscript, we examine a spherical symmetric
vector potential in the form of
\end{subequations}
\begin{equation}
\overrightarrow{A}=\frac{A_{r}(r)}{r}\overrightarrow{r}.
\end{equation}%
Then, by using the properties of vector spherical harmonics \cite%
{AITTAHAR1995307, BARRETT1995311, doi:10.1063/1.3494292}, we obtain the
following radial differential equations:

\begin{eqnarray}
\sqrt{1-\alpha r^{2}}\zeta _{J}\left( \frac{d}{dr}-\frac{J+1}{r}-\frac{A_{r}%
}{\sqrt{1-\alpha r^{2}}}\right) F_{0} &=&-mH_{+1},  \label{eq01} \\
\sqrt{1-\alpha r^{2}}\xi _{J}\left( \frac{d}{dr}+\frac{J}{r}-\frac{A_{r}}{%
\sqrt{1-\alpha r^{2}}}\right) F_{0} &=&-mH_{-1},  \label{eq02} \\
\sqrt{1-\alpha r^{2}}\left[ \zeta _{J}\left( \frac{d}{dr}+\frac{J+1}{r}-%
\frac{A_{r}}{\sqrt{1-\alpha r^{2}}}\right) F_{+1}+\xi _{J}\left( \frac{d}{dr}%
-\frac{J}{r}-\frac{A_{r}}{\sqrt{1-\alpha r^{2}}}\right) F_{-1}\right]
&=&-mH_{0},  \label{eq03} \\
\sqrt{1-\alpha r^{2}}\left[ -\xi _{J}\left( \frac{d}{dr}+\frac{J+1}{r}+\frac{%
A_{r}}{\sqrt{1-\alpha r^{2}}}\right) G_{+1}+\zeta _{J}\left( \frac{d}{dr}-%
\frac{J}{r}+\frac{A_{r}}{\sqrt{1-\alpha r^{2}}}\right) G_{-1}\right]
&=&m\varphi ,  \label{eq04} \\
-\sqrt{1-\alpha r^{2}}\zeta _{J}\left( \frac{d}{dr}-\frac{J+1}{r}+\frac{A_{r}%
}{\sqrt{1-\alpha r^{2}}}\right) H_{0}+\left( E-iA_{0}\right) G_{+1}
&=&mF_{+1},  \label{eq05} \\
-\sqrt{1-\alpha r^{2}}\xi _{J}\left( \frac{d}{dr}+\frac{J}{r}+\frac{A_{r}}{%
\sqrt{1-\alpha r^{2}}}\right) H_{0}+\left( E-iA_{0}\right) G_{-1} &=&mF_{-1},
\label{eq06} \\
-\sqrt{1-\alpha r^{2}}\left[ \zeta _{J}\left( \frac{d}{dr}+\frac{J+1}{r}+%
\frac{A_{r}}{\sqrt{1-\alpha r^{2}}}\right) H_{+1}+\xi _{J}\left( \frac{d}{dr}%
-\frac{J}{r}+\frac{A_{r}}{\sqrt{1-\alpha r^{2}}}\right) H_{-1}\right]
+\left( E-iA_{0}\right) G_{0} &=&mF_{0},  \label{eq07} \\
\left( E+iA_{0}\right) F_{0} &=&mG_{0},  \label{eq08} \\
-\sqrt{1-\alpha r^{2}}\xi _{J}\left( \frac{d}{dr}-\frac{J-1}{r}-\frac{A_{r}}{%
\sqrt{1-\alpha r^{2}}}\right) \varphi +\left( E+iA_{0}\right) F_{+1}
&=&mG_{+1},  \label{eq09} \\
\sqrt{1-\alpha r^{2}}\zeta _{J}\left( \frac{d}{dr}+\frac{J}{r}-\frac{A_{r}}{%
\sqrt{1-\alpha r^{2}}}\right) \varphi +\left( E+iA_{0}\right) F_{-1}
&=&mG_{-1}.  \label{eq10}
\end{eqnarray}%
where 
\begin{equation*}
\xi _{J}=\sqrt{\frac{J+1}{2J+1}},\qquad \zeta _{J}=\sqrt{\frac{J}{2J+1}}.
\end{equation*}%
Nedjadi \emph{et al.}, in \cite{doi:10.1063/1.530801}, presented a procedure
to decouple ten-coupled differential equations into two classes of coupled
differential equation sets by taking the parity into account. We follow
their procedure and consider Eqs. \eqref{eq01}, \eqref{eq02}, \eqref{eq07},
and \eqref{eq08} for the natural parity states which relates $F_{0}$, $G_{0}$%
, $H_{+1}$ and $H_{-1}$ functions. We take $\varphi $, $H_{0}$,$F_{+1}$, $%
F_{-1}$, $G_{+1}$, and $G_{-1}$ functions as zero. On the other hand, for
the unnatural parity states, we examine Eqs. \eqref{eq03}, \eqref{eq04}, %
\eqref{eq05}, \eqref{eq06}, \eqref{eq09}, and \eqref{eq10} that associate $%
F_{+1}$, $F_{-1}$, $G_{+1}$, $G_{-1}$, $H_{0}$, and $\varphi $ functions,
while we assume $F_{0}$, $G_{0}$, $H_{+1}$ and $H_{-1}$, thus, Eqs. %
\eqref{eq01}, \eqref{eq02}, \eqref{eq07}, and \eqref{eq08} are zero.

\subsection{$\left(-1\right)^{J}$ parity states}

\begin{figure}[hbtp]
 \centering
 \includegraphics[scale=1]{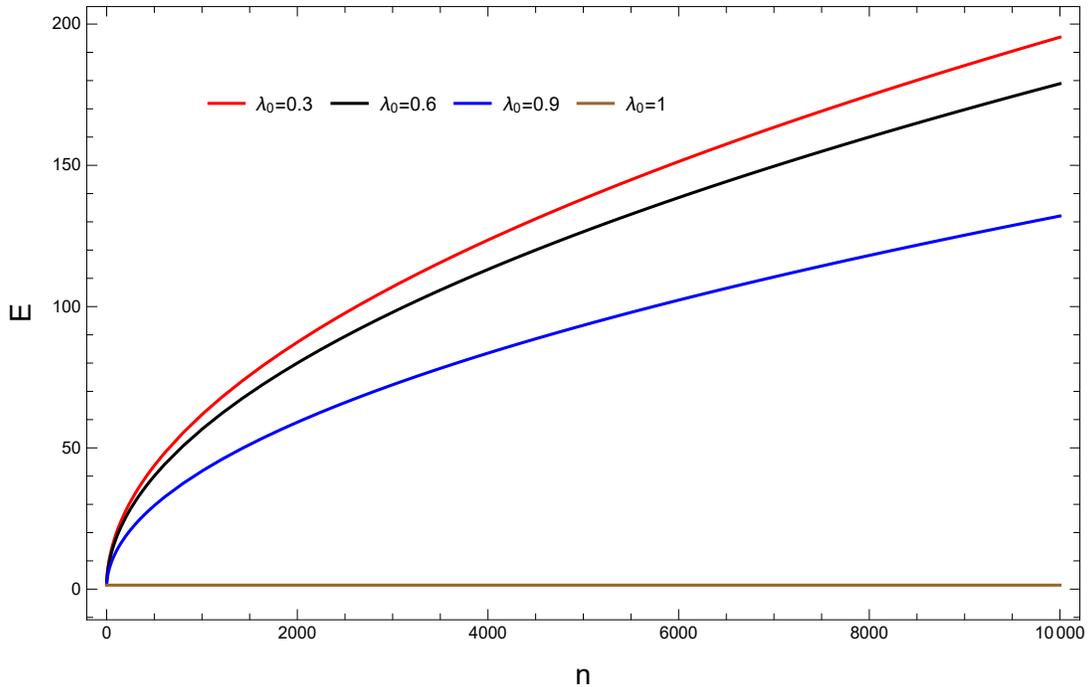}
 \caption{Energy levels versus the quantum number, $n$, for different values of $\lambda_{0}$.} \label{fig1}
 \end{figure}

In this subsection we obtain the energy eigenvalue function for the natural
parity states. From Eqs. \eqref{eq01}, \eqref{eq02}, and \eqref{eq08} we
find $H_{+1}$, $H_{-1}$, $G_{0}$ in terms of $F_{0}$. Then, we employ them
in Eq. \eqref{eq07}. After a little algebra we find
\begin{equation}
\bigg[\left( 1-\alpha r^{2}\right) \left( \frac{d^{2}}{dr^{2}}-\frac{J\left(
J+1\right) }{r^{2}}-\frac{d}{dr}\frac{A_{r}}{\sqrt{1-\alpha r^{2}}}-\frac{%
A_{r}^{2}}{1-\alpha r^{2}}\right) -\alpha r\left( \frac{d}{dr}-\frac{A_{r}}{%
\sqrt{1-\alpha r^{2}}}\right) +\left( E^{2}+A_{0}^{2}-m^{2}\right) \bigg]%
F_{0}=0.\,\,\,\, \label{eqf02}
\end{equation}%
Then, we consider a vector potential energy with the following components:
\begin{eqnarray}
A_{0} &=&\lambda _{0}\frac{r}{\sqrt{1-\alpha r^{2}}},  \label{A0} \\
A_{r} &=&\lambda _{r}\frac{r}{\sqrt{1-\alpha r^{2}}},  \label{Ar}
\end{eqnarray}%
where $\lambda _{0}$ and $\lambda _{r}$ are coupling constants that obey $\lambda _{r}\geq \lambda _{0}$ condition. By
employing the vector potential components in Eq. \eqref{eqf02}, we arrive at
\begin{equation}
\bigg[\left( \sqrt{1-\alpha r^{2}}\frac{d}{dr}\right) ^{2}-\left( 1-\alpha
r^{2}\right) \frac{J\left( J+1\right) }{r^{2}}-\frac{\lambda _{r}}{1-\alpha
r^{2}}-\frac{\left( \lambda _{r}^{2}-\lambda _{0}^{2}\right) r^{2}}{1-\alpha
r^{2}}+E^{2}-m^{2}\bigg]F_{0}=0.
\end{equation}%
Next, we introduce a new variable $\rho $ via the coordinate transformation $%
\rho =\alpha r^{2}$. We obtain
\begin{equation}
\left[ \left( 1-\rho \right) \rho \frac{d^{2}}{d\rho ^{2}}+\left( \frac{1}{2}%
-\rho \right) \frac{d}{d\rho }-\frac{J\left( J+1\right) }{4\rho }-\frac{1}{4}%
\frac{\left( \frac{\lambda _{r}^{2}}{\alpha ^{2}}+\frac{\lambda _{r}}{\alpha
}-\frac{\lambda _{0}^{2}}{\alpha ^{2}}\right) }{1-\rho }+\frac{E^{2}-m^{2}}{%
4\alpha }+\frac{J\left( J+1\right) }{4}+\frac{\lambda _{r}^{2}-\lambda
_{0}^{2}}{4\alpha ^{2}}\right] F_{0}=0.  \label{eqf055}
\end{equation}%
For the general solution, we follow an Ansatz as
\begin{equation}
F_{0}=\rho ^{a}\left( 1-\rho \right) ^{b}\varXi(\rho ).    \label{coz}
\end{equation}%
Then, we reach
\begin{equation}
\Bigg[\left( 1-\rho \right) \rho \frac{d^{2}}{d\rho ^{2}}+\left( \frac{1}{2}%
+2a-\left( 1+2a+2b\right) \rho \right) \frac{d}{d\rho }+\frac{v_{1}}{\rho }+%
\frac{v_{2}}{1-\rho }-u\Bigg]\varXi(\rho )=0,  \label{eqf05}
\end{equation}%
where
\begin{eqnarray}
v_{1} &=&a\left( a-\frac{1}{2}\right) -\frac{J\left( J+1\right) }{4}, \\
v_{2} &=&b\left( b-\frac{1}{2}\right) -\frac{1}{4}\left( \frac{\lambda
_{r}^{2}}{\alpha ^{2}}+\frac{\lambda _{r}}{\alpha }-\frac{\lambda _{0}^{2}}{%
\alpha ^{2}}\right) , \\
u &=&\left( a+b\right) ^{2}-\frac{E^{2}-m^{2}}{4\alpha }-\frac{J\left(
J+1\right) }{4}-\frac{\lambda _{r}^{2}-\lambda _{0}^{2}}{4\alpha ^{2}}.
\end{eqnarray}%
For the roots
\begin{subequations}
\begin{eqnarray}
a &=&\frac{J+1}{2}, \\
b &=&\frac{1}{4}+\frac{1}{4}\sqrt{1+4\bigg[\frac{\lambda _{r}}{\alpha }%
\left( \frac{\lambda _{r}}{\alpha }+1\right) -\frac{\lambda _{0}^{2}}{\alpha
^{2}}\bigg]},
\end{eqnarray}%
$v_{1}$ and $v_{2}$ vanish. Therefore, Eq. \eqref{eqf05} reduces to the
hypergeometric differential equation
\end{subequations}
\begin{equation*}
\left[ \rho \left( 1-\rho \right) \frac{d^{2}}{d\rho ^{2}}+\Big(C-\left(
1+A+B\right) \rho \Big)\frac{d}{d\rho }-AB\right] \varXi(\rho )=0,
\end{equation*}%
where
\begin{equation*}
\varXi=N_{1}\times {}_{2}F_{1}\left( A;B;C;\rho \right).
\end{equation*}%
Here, $N_{1}$ is the normalization constant, and
\begin{subequations}
\begin{eqnarray}
A &=&a+b+\sqrt{\frac{E^{2}-m^{2}}{4\alpha }+\frac{J\left( J+1\right) }{4}+%
\frac{\lambda _{r}^{2}-\lambda _{0}^{2}}{4\alpha ^{2}}}, \\
B &=&a+b-\sqrt{\frac{E^{2}-m^{2}}{4\alpha }+\frac{J\left( J+1\right) }{4}+%
\frac{\lambda _{r}^{2}-\lambda _{0}^{2}}{4\alpha ^{2}}}, \\
C &=&\frac{1}{2}+2a.
\end{eqnarray}%
\end{subequations}
We want to emphasize that to obtain a nonsingular solution, we do not consider the second solution of the hypergeometric differential equation by equating its normalization constant to zero. After that for the quantization, we use the well-known condition
\begin{equation}
B=-n \label{quan}
\end{equation}%
which yields to
\begin{equation}
E_{n;J}=\pm \sqrt{m^{2}+4\alpha \left( n+\frac{2J+3}{4}+\frac{1}{4}\sqrt{1+4%
\bigg[\frac{\lambda _{r}}{\alpha }\left( \frac{\lambda _{r}}{\alpha }%
+1\right) -\frac{\lambda _{0}^{2}}{\alpha ^{2}}\bigg]}\right) ^{2}-\alpha
J\left( J+1\right) -\frac{\left( \lambda _{r}^{2}-\lambda _{0}^{2}\right) }{%
\alpha }}. \label{En1}
\end{equation}
We see that the energy eigenvalue expression contains an additional correction term which depends on the deformation $\alpha $.  It is worth noting that the presence of a correction term proportional to $n^{2}$ indicates the appearance of hard confinement due to the deformation. This is similar to the energy eigenvalue function of a particle in a square well potential whose boundaries are placed at $\pm \frac{\pi}{2\sqrt{\alpha }}$.  The second correction term is proportional to $J\left( J+1\right) $, so, it mimics a kind of rotational energy and removes the degeneracy of the usual spectrum according to $J$ number. In addition, in the limit of $\alpha \rightarrow 0$, the energy level for the spin-one DKPe with a nonminimal vector interaction reduces to
\begin{equation}
E_{n;J}=\sqrt{m^{2}+\lambda _{r}+\left( 4n+J+3\right) \sqrt{\lambda
_{r}^{2}-\lambda _{0}^{2}}}, \label{alpha0limit}
\end{equation}
which is the same result of ordinary case \cite{Castro_et_al_2014}. We demonstrate these results graphically by assigning some numerical values to the
deformation and nonminimal vector coupling parameter $\lambda _{0}$. Note that in all graphs of this manuscript we assume $m=1$ and $\lambda _{r}=1$. In fig.~\ref{fig1}, for
$\alpha=J=0$, we present the behavior of the energy eigenvalue function versus $n$ with four different values of $\lambda _{0}$. We see a constant energy value for $\lambda _{0}=\lambda _{r}$ as foreseen in Eq. \eqref{alpha0limit}. On the other hand, we observe that for decreasing values of A, the energy increases faster for small n quantum numbers. We observe that the increments of these increases gradually slow down as the quantum numbers become larger.

\begin{figure}[hbtp]
 \centering
 \includegraphics[scale=1]{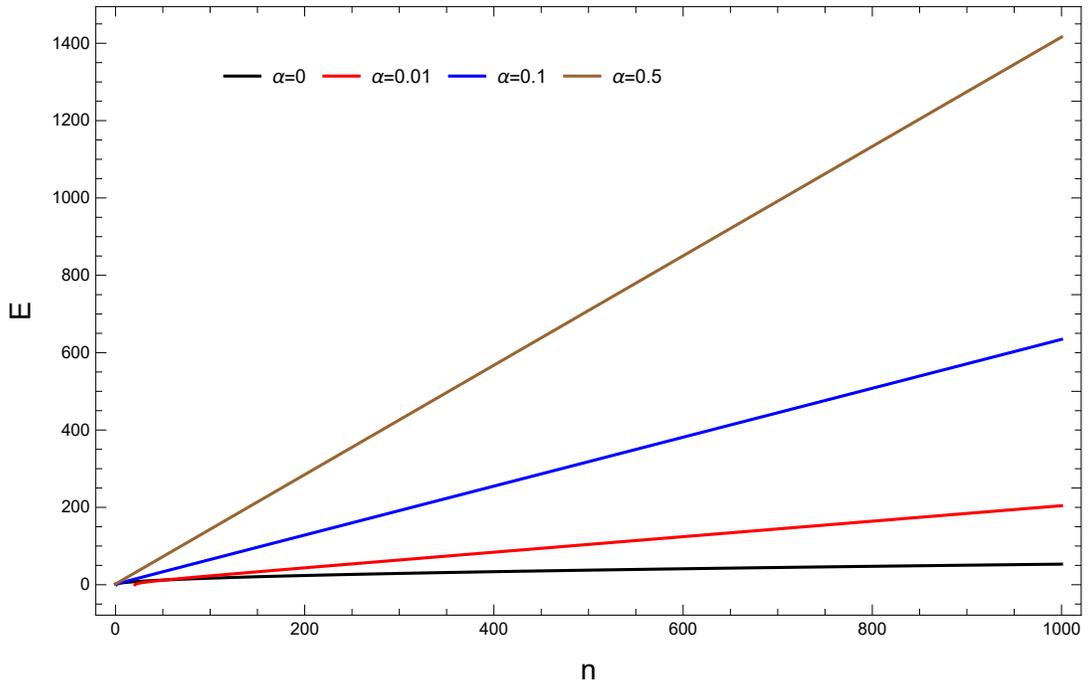}
 \caption{Energy levels versus the quantum number $n$ for different values of the
 deformation parameters $\alpha$}  \label{fig2}
 \end{figure}

Next, we assign a fixed value for the vector coupling parameter, $\lambda _{0}=0.5$, with $J=0$ and examine the behaviour of the energy eigenvalue function versus $n$ with four different values of the deformation parameter in fig.~\ref{fig2}. When $\alpha$ is equal to zero, we observe an increase in the energy function as predicted in Eq.~\eqref{alpha0limit}. For non-zero alpha values, we see that the increase in energy function varies linearly in accordance with Eq.~\eqref{En1} with respect to the quantum number $n$. The increase is greater in larger deformation parameters.

\begin{figure}[hbtp]
 \centering
 \includegraphics[scale=1]{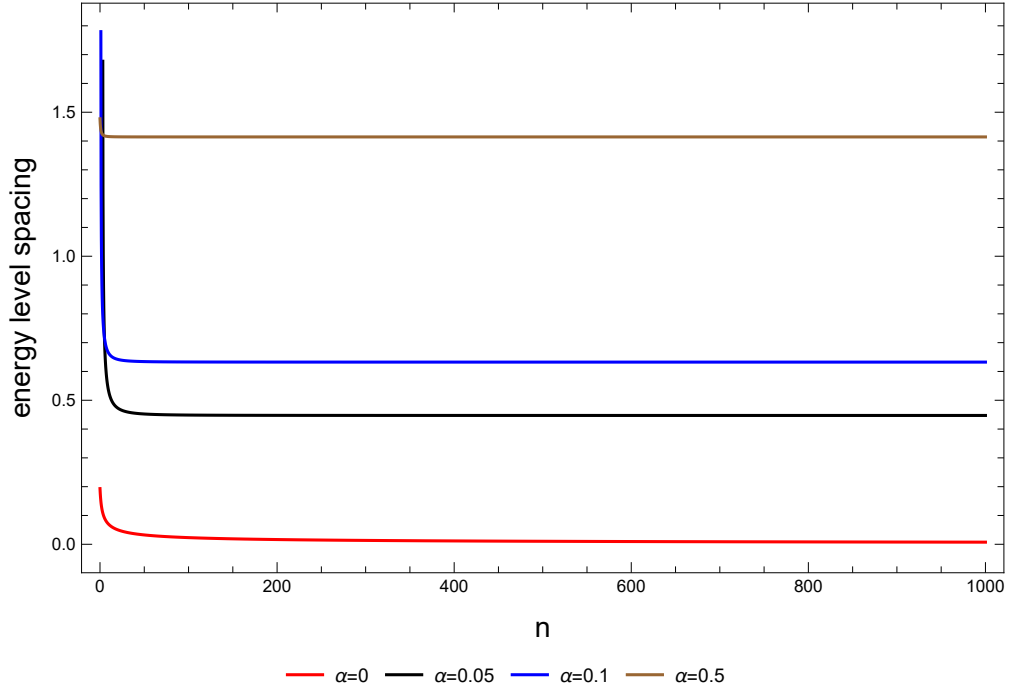}
 \caption{ Energy levels spacing versus the quantum number $n$ for different values of the
 deformation parameters $\alpha$.} \label{fig3}
 \end{figure}

Another interesting property arises for the
energy level spacing which is defined by $\Delta E_{n;J}=E_{n+1;J}-E_{n;J}$. We observe that for large $n$,%
\begin{equation}
\lim_{n\rightarrow \infty }\left\vert \Delta E_{n;J}\right\vert =2\sqrt{%
\alpha}, \label{Ens}
\end{equation}
It is worth noting that the energy spacing tends to zero in the absence of the MUM. For the graphical illustration, we take $\lambda_0=0.5$ and  plot the energy level spacing versus quantum number $n$ for $J=0$ with different values of deformation parameter in fig.~\ref{fig3}.

We observe that for small $n$, the changes in between the energy levels are not constant. For higher quantum numbers, the energy level spacings become constant and it is proportional to the deformation parameter as predicted in Eq.~\eqref{Ens}.

\subsection{ $\left(-1\right)^{J+1}$ parity states}

In this subsection we investigate the unnatural parity states and derive the
energy eigenvalue function. In the general case, where $A_{0}$ and $A_{r}$
have non zero values as given in Eqs. \eqref{A0} and \eqref{Ar}. Castro
\emph{et al.} \cite{Castro_et_al_2014} stated that a decoupling process can
not be executed successfully for the Eqs. \eqref{eq03}, \eqref{eq04}, %
\eqref{eq05}, \eqref{eq06}, \eqref{eq09}, and \eqref{eq10}. Instead, for $%
\lambda _{0}=0$, thus, $A_{0}=0$, Eqs. \eqref{eq05}, \eqref{eq06}, %
\eqref{eq09}, and \eqref{eq10} reduce to the following forms, respectively.
\begin{eqnarray}
-\sqrt{1-\alpha r^{2}}\zeta _{J}\left( \frac{d}{dr}-\frac{J+1}{r}+\frac{A_{r}%
}{\sqrt{1-\alpha r^{2}}}\right) H_{0} &=&mF_{+1}-EG_{+1},  \label{eq051} \\
-\sqrt{1-\alpha r^{2}}\xi _{J}\left( \frac{d}{dr}+\frac{J}{r}+\frac{A_{r}}{%
\sqrt{1-\alpha r^{2}}}\right) H_{0} &=&mF_{-1}-EG_{-1},  \label{eq061} \\
-\sqrt{1-\alpha r^{2}}\xi _{J}\left( \frac{d}{dr}-\frac{J-1}{r}-\frac{A_{r}}{%
\sqrt{1-\alpha r^{2}}}\right) \varphi &=&mG_{+1}-EF_{+1},  \label{eq091} \\
\sqrt{1-\alpha r^{2}}\zeta _{J}\left( \frac{d}{dr}+\frac{J}{r}-\frac{A_{r}}{%
\sqrt{1-\alpha r^{2}}}\right) \varphi &=&mg_{-1}-EF_{-1}.  \label{eq101}
\end{eqnarray}%
After a little algebra, we express these four equations in the form of
\begin{equation}
\left(
\begin{array}{c}
F_{+1} \\
G_{+1}%
\end{array}%
\right) =\frac{\sqrt{1-\alpha r^{2}}}{E^{2}-m^{2}}\left(
\begin{array}{cc}
E\xi _{J}\left( \frac{d}{dr}-\frac{J+1}{r}-\frac{A_{r}}{\sqrt{1-\alpha r^{2}}%
}\right) & m\zeta _{J}\left( \frac{d}{dr}-\frac{J+1}{r}+\frac{A_{r}}{\sqrt{%
1-\alpha r^{2}}}\right) \\
m\xi _{J}\left( \frac{d}{dr}-\frac{J+1}{r}-\frac{A_{r}}{\sqrt{1-\alpha r^{2}}%
}\right) & E\zeta _{J}\left( \frac{d}{dr}-\frac{J+1}{r}+\frac{A_{r}}{\sqrt{%
1-\alpha r^{2}}}\right)%
\end{array}%
\right) \left(
\begin{array}{c}
\varphi \\
H_{0}%
\end{array}%
\right)  \label{bcl1}
\end{equation}%
and
\begin{equation}
\left(
\begin{array}{c}
f_{-1} \\
g_{-1}%
\end{array}%
\right) =\frac{\sqrt{1-\alpha r^{2}}}{E^{2}-m^{2}}\left(
\begin{array}{cc}
-E\zeta _{J}\left( \frac{d}{dr}+\frac{J}{r}-\frac{A_{r}}{\sqrt{1-\alpha r^{2}%
}}\right) & m\xi _{J}\left( \frac{d}{dr}+\frac{J}{r}+\frac{A_{r}}{\sqrt{%
1-\alpha r^{2}}}\right) \\
-m\zeta _{J}\left( \frac{d}{dr}+\frac{J}{r}-\frac{A_{r}}{\sqrt{1-\alpha r^{2}%
}}\right) & E\xi _{J}\left( \frac{d}{dr}+\frac{J}{r}+\frac{A_{r}}{\sqrt{%
1-\alpha r^{2}}}\right)%
\end{array}%
\right) \left(
\begin{array}{c}
\varphi \\
h_{0}%
\end{array}%
\right)  \label{bcl2}
\end{equation}%
We take $g_{+1}$ and $g_{-1}$ from Eqs. \eqref{bcl1}, \eqref{bcl2} and
employ in Eq. \eqref{eq04}. We find
\begin{equation}
\bigg[\left( 1-\alpha r^{2}\right) \frac{d^{2}}{dr^{2}}-\alpha r\frac{d}{dr}-%
\frac{J\left( J+1\right) }{r^{2}}-\frac{\alpha \lambda _{r}\left( \frac{%
\lambda _{r}}{\alpha }-1\right) r^{2}}{\left( 1-\alpha r^{2}\right) }+\bar{k}%
_{1}\bigg]H_{0}+\frac{\alpha E}{m}\sqrt{J\left( J+1\right) }\varphi =0,
\label{eqf041}
\end{equation}%
where $\bar{k}_{1}=E^{2}-m^{2}+\lambda _{r}+\alpha J\left( J+1\right) $.
Alike, we draw $f_{+1}$ and $f_{-1}$ from Eqs. \eqref{bcl1}, \eqref{bcl2}
and use in Eq. \eqref{eq03}. We get
\begin{equation}
\bigg[\left( 1-\alpha r^{2}\right) \frac{d^{2}}{dr^{2}}-\alpha r\frac{d}{dr}-%
\frac{J\left( J+1\right) }{r^{2}}-\frac{\alpha \lambda _{r}\left( \frac{%
\lambda _{r}}{\alpha }-1\right) r^{2}}{\left( 1-\alpha r^{2}\right) }+\bar{k}%
_{2}\bigg]\varphi +\frac{\alpha E}{m}\sqrt{J\left( J+1\right) }H_{0}=0,
\label{eqf031}
\end{equation}%
while $\bar{k}_{2}=E^{2}-m^{2}-\left( 3\lambda _{r}-\alpha \right) +\alpha
J\left( J+1\right) $. For simplicity, we set $J=0$, and examine a particular
solution among the general solution. It is worth noting that under this
choice, $H_{0}$ and $\varphi $ decouple from each each others in Eqs. %
\eqref{eqf031} and \eqref{eqf041} as
\begin{eqnarray}
\bigg[\left( 1-\alpha r^{2}\right) \frac{d^{2}}{dr^{2}}-\alpha r\frac{d}{dr}-%
\frac{\alpha \lambda _{r}\left( \frac{\lambda _{r}}{\alpha }-1\right) r^{2}}{%
\left( 1-\alpha r^{2}\right) }+E^{2}-m^{2}-\left( 3\lambda _{r}-\alpha
\right) \bigg]\varphi &=&0,  \label{eqf032} \\
\bigg[\left( 1-\alpha r^{2}\right) \frac{d^{2}}{dr^{2}}-\alpha r\frac{d}{dr}-%
\frac{\alpha \lambda _{r}\left( \frac{\lambda _{r}}{\alpha }-1\right) r^{2}}{%
\left( 1-\alpha r^{2}\right) }+E^{2}-m^{2}+\lambda _{r}\bigg]H_{0} &=&0.
\label{eqf042}
\end{eqnarray}%
At the next step, we introduce a variable change $\rho =\alpha r^{2}$. We
obtain
\begin{eqnarray}
\bigg[(1-\rho )\rho \frac{d^{2}}{d\rho ^{2}}+\left( \frac{1}{2}-\rho \right)
\frac{d}{d\rho }+\frac{E^{2}-m^{2}}{4\alpha }+\frac{1}{4}-\frac{\frac{%
\lambda _{r}}{\alpha }\left( \frac{\lambda _{r}}{\alpha }+1\right) }{%
4(1-\rho )}+\frac{\lambda _{r}}{4\alpha }\left( \frac{\lambda _{r}}{\alpha }%
-2\right) \bigg]\varphi &=&0,  \label{eqf033} \\
\bigg[(1-\rho )\rho \frac{d^{2}}{d\rho ^{2}}+\left( \frac{1}{2}-\rho \right)
\frac{d}{d\rho }+\frac{E^{2}-m^{2}}{4\alpha }+\frac{\lambda _{r}^{2}}{%
4\alpha ^{2}}-\frac{\frac{\lambda _{r}}{\alpha }\left( \frac{\lambda _{r}}{%
\alpha }-1\right) }{4(1-\rho )}\bigg]H_{0} &=&0.  \label{eqf043}
\end{eqnarray}
Since Eqs. \eqref{eqf033} and \eqref{eqf043} are similar to Eq. \eqref{eqf055}, they can be
solved exactly in the same manner. We follow the recipe written in between Eqs. \eqref{coz}  and \eqref{quan} and obtain the
energy spectra as%
\begin{eqnarray}
E_{\varphi } &=&\pm \sqrt{m^{2}+4\lambda _{r}+4\alpha \left( n+\frac{1}{2}%
\right) \left( n+\frac{3}{2}+\frac{\lambda _{r}}{\alpha }\right) }, \\
E_{H_{0}} &=&\pm \sqrt{m^{2}+4\alpha \left( n+\frac{1}{2}\right) \left( n+%
\frac{1}{2}+\frac{\lambda _{r}}{\alpha }\right) }.
\end{eqnarray}

\begin{figure}[hbtp]
 \centering
 \includegraphics[scale=1]{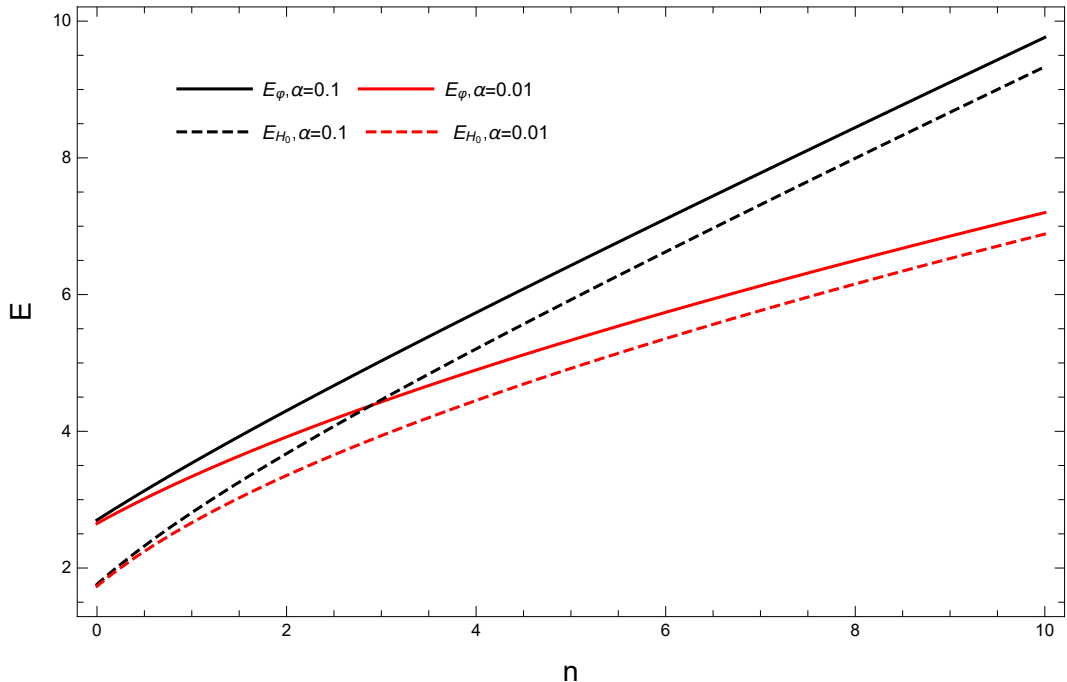}
 \caption{Variation of DKP energies with $n$ for different values of the
 deformation parameters $\alpha$.} \label{fig4}
 \end{figure}

Finally, we take $J=0$ and $\lambda_0=0.5$ and plot the energy functions $E_{\varphi }$ and $E_{H_{0}} $ versus $n$ for two nonzero deformation parameter values in fig.~\ref{fig4}. We observe that for all quantum numbers $E_{\varphi }$ is greater than $E_{H_{0}}$. The change of the deformation parameter in small quantum numbers does not cause much difference in the energy values. As the quantum numbers grow, the change of the deformation parameter has a greater effect on the energy values. For $E_{\varphi }$ and $E_{H_{0}}$, these effects are similar.

\section{Conclusion} \label{sec4}
In this paper, we discussed the various consequences of considering a non-minimal vector interaction in the presence of minimal uncertainty in momentum in the Duffin-Kemmer-Petiau (DKP) formalism. We obtained eigenfunctions in terms of the hypergeometric function, analytically. We exposed an explicit calculation of the energy eigenvalue function for the bound states of the spin-one DKP equation in three-dimension spaces by using the quantization condition. Since the energy eigenfunctions depend on the nonminimal coupling constants and the deformation parameters, we revealed the effects of them on the energy values analytically. We strengthened our findings by presenting these effects in several figures. Moreover, we found that the energy eigenvalues depend on the quantum number $n^{2}$ like square well problem.  For large $n$, we found that the energy level spacings become a constant which is proportional to the deformation parameter. Our finding predicted a discontinuity in the energy levels.

\section*{Acknowledgment}
One of the author, B.C. L\" utf\"uo\u{g}lu, was partially supported by the Turkish Science and Research Council (T\"{U}B\.{I}TAK).

\bibliography{BCLmain}




\end{document}